\documentclass[twocolumn,showpacs,preprintnumbers,amsmath,amssymb]{revtex4}

\usepackage{graphicx}
\usepackage{dcolumn}
\usepackage{bm}
\usepackage{amsmath}

\begin{document}
\title{A theory of the three-pulse electric-dipole echo in glasses in a magnetic field}

\author{Y.\,M. Beltukov}
\email{ybetukov@gmail.com}
\author{D.\,A. Parshin}
\affiliation{Saint Petersburg State Technical University, 195251 Saint Petersburg, Russia}

\date{\today}

\begin{abstract}
We extended existing theory of the two-pulse electric-dipole echo in glasses in a magnetic field to the three-pulse echo. As is well known, at low temperatures two-level systems (TLS's) are responsible for the echo phenomenon in glasses. Using a diagram technique in the framework of perturbation theory we derived a simple formula for the three-pulse echo amplitude. As in the case of two-pulse echo the magnetic field dependence of the tree-pulse echo amplitude in glasses is related to quadrupole electric moments of TLS's non-spherical nuclei and/or dipole-dipole interaction of their nuclear spins. These two mechanisms are responsible for the fine level splitting of TLS. As a result TLS transforms to multi-level system with the fine level splitting depending on a magnetic field. Due to existence in the theory the additional parameter $T$ --- the time interval between the second and the third pulses we have more reach spectrum of echo oscillations in a magnetic field in comparison with the case of the two-pulse echo.
\end{abstract}


\maketitle

\section{Introduction}

It is well known that glasses at low temperatures (below 1 K) display a number of universal properties which are drastically different from the properties of similar crystals. These properties are almost independent of the chemical composition of a glass and mainly due to the disordered atomic structure of an amorphous solid; more exactly, due to the absence of a long-range order in the glass \cite{AGVF}.

According to the standard theory of Anderson, Halperin, Varma, and Phillips \cite{AGVF}, all these universal properties are associated with the existence of two-level systems (TLS's) in glasses. TLS consist of atoms or groups of atoms that may tunnel between two minima in a double-well potential. However, the microscopic nature of the TLS's in the majority of glasses remains unknown.

One of such universal phenomena is an electric-dipole echo \cite{AGVF}. Depending on the sequence of applied high frequency electric pulses one usually discriminates between two-pulse and three-pulse echo. The essence of the two-pulse echo is the following. When a glass is subjected to two short electromagnetic pulses at
a frequency of about 1 GHz separated by a time interval $\tau$, a response signal at the same frequency may be detected in the dipole moment of the system at time
$\tau$ after the second pulse. In the three-pulse echo we have a sequence of three pumping electromagnetic  pulses in the moments $0$, $\tau$ and $\tau + T$ correspondingly. A coherent echo signal in this case appears at the time $T+2\tau$.

A very interesting echo phenomenon was discovered recently in 2002 \cite{albaze}. The amplitude of the two-pulse dipole echo in some nonmagnetic glasses exhibited a strong non-monotonic (oscillating) dependence on the magnetic field even in weak fields (about 10 mT). Wurger, Fleischmann, and Enss \cite{quadr} suggested that this unusual  magnetic field effect is caused by the presence of non-spherical tunneling nuclei with the electric quadrupole moment (or, equivalently, the nuclear spin $J\ge 1$) in the glass. The interaction of the quadrupole moment with gradient of static microscopic electric field in two wells of double-well potential creates a fine splitting of energy levels in TLS transforming the TLS to multi-level system.

This idea agrees very well with the experimental data, in particular, with the results of the recent measurements of the two-pulse dipole echo in glycerol (C$_3$H$_8$O$_3$) \cite{glyc}. In these experiments, it was shown that the replacement of hydrogen, which has nuclear spin $J= 1/2$ and, consequently, a zero quadrupole moment, by deuterium ($J = 1$ and a nonzero quadrupole moment) results in a more than an order of magnitude enhancement of the magnetic-field dependence of the echo.

The magnetic-field dependence of the two-pulse echo amplitude in glasses with nuclear quadrupole moments was theoretically studied by Wurger, Fleischmann, and Enss \cite{quadr,wurger} and by one of us \cite{diagram, ShumPar1}. However the weak magnetic field non monotonic dependence of echo in non-deuterated glycerol could not be explained in such a way. Neither of nuclei forming glycerol glass C$_3$H$_8$O$_3$ possess a nuclear quadrupole moment.

To explain this interesting phenomenon in non-deuterated glycerol Bazrafshan et al. \cite{JOP} suggested that the magnetic field dependence of the two-pulse echo amplitude in this case may be caused by the dipole-dipole interaction of the nuclear magnetic moments of hydrogen atoms. This interaction (along with Zeeman interaction) can also create a fine structure of two levels in TLS, which depends on the applied magnetic field.

In the papers \cite{JOP,ltc08} the two-pulse echo amplitude was numerically calculated in deuterated glycerol C$_3$D$_5$H$_3$O$_3$ (taking into account all of the spins of hydrogen and the quadrupole moments of deuterium nuclei), assuming that the tunneling motion in the two level system is the rotation of the glycerol molecule as a whole. Results of these calculations were in a good agreement with experimental data. Later in \cite{ShumPar2} an analytical theory of the magnetic field dependence of the two-pulse echo amplitude in glasses with dipole-dipole interaction of nuclear spins was developed. Without any assumptions about microscopic tunneling mechanism of hydrogen atoms in glycerol and without fitting parameters it also shows a very good agreement with experimental data.

The purpose of this paper is to build a theory of the three-pulse echo in glasses in a magnetic field to stimulate further experimental investigations of this interesting phenomenon. We consider here two mechanisms of magnetic field effects, namely quadrupole electric moments of TLS's nuclei and dipole-dipole magnetic interaction of their nuclear spins.

\section{Three-pulse Echo in an Ensemble of TLS}

\begin{figure}
    \centering
    \includegraphics[width=0.45\textwidth]{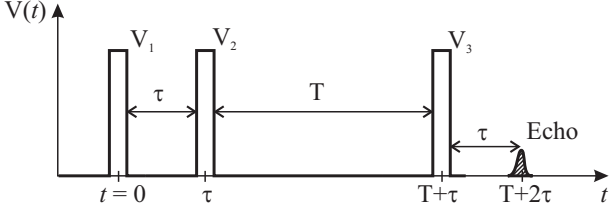}
    \caption{\label{fig:Echo2} Electric pulses and the three-pulse echo in the TLS.}
\end{figure}
We start with a simple example of a three-pulse echo in an ensemble of TLS. There are three exciting electromagnetic  pulses with amplitudes ${\bf F}_1$, ${\bf F}_2$, and ${\bf F}_3$ correspondingly. The time interval between the first and the second pulse is $\tau$, between the second and the third pulse --- is $T$ (Fig.~\ref{fig:Echo2}). We suppose that the duration of any pulse is much less than the time delays between the pulses.

First we consider one TLS. The wave function $\Psi$ of the two level-system is a linear combination of wave functions $\varphi_1$ and $\varphi_2$ for each level
\begin{equation}
    \Psi = C_1\varphi_1+C_2\varphi_2, \quad |C_1|^2 + |C_2|^2 = 1 .
\end{equation}
Initial conditions (just before the first pulse) are
\begin{equation}
    C_1 = 1, \quad C_2 = 0.
\end{equation}
Then in the electric field the time variation of the amplitudes $C_1$ and $C_2$ obeys the equations
\begin{equation}
    i\hbar\frac{dC_1}{dt} = E_1C_1+V(t)C_2,   \quad
    i\hbar\frac{dC_2}{dt} = E_2C_2+V(t)C_1.
\end{equation}
For the off-diagonal transition matrix element $V(t)$ we have a following expression during the electric pulse
\begin{equation}
V(t)=V_{1,2,3}\cos\omega t , \quad
V_{1,2,3}=({\bf F}_{1,2,3}
\cdot{\bf m})\frac{\Delta_0}{E} .
\label{eq:e1a}
\end{equation}
Here ${\bf F}_{1,2,3}$ is the electric field amplitude of the first, second and the third electric pulse, respectively, ${\bf m}$ is the dipole moment of the TLS, $\Delta_0$ is the tunneling splitting, $E=E_2-E_1$ is the TLS energy, and $E_1$ and $E_2$ are the energies of the ground and the excited states of the TLS, respectively. The electric field frequency is assumed as $\hbar\omega\approx E$. To simplify our equations we will put $\hbar=1$.

After the action of the 1st electric pulse $C_2$ acquires in the first approximation a finite value proportional to amplitude $V_1$ which is assumed to be small. During the time interval between the first and the second pulses
($0<t<\tau$) we have
\begin{equation}\label{C1C2between12}
    C_1 \propto 1\cdot e^{-iE_1t},  \quad
    C_2 \propto  -iV_1e^{-iE_2t},
\end{equation}
where we have neglected terms proportional to $V_1^2$. In a similar way after the action of the second pulse we have between the second and the third pulses
\begin{eqnarray}\label{C1C2between23}
    C_1 &\propto& 1\cdot e^{-iE_1t}-V_1V_2e^{-iE_2\tau}e^{-iE_1(t-\tau)},   \nonumber\\
    C_2 &\propto& -iV_1e^{-iE_2t}-iV_2e^{-iE_1\tau}e^{-iE_2(t-\tau)}.
\end{eqnarray}
Then finally after the third pulse we get
\begin{multline}\label{C1after3}
    C_1 \propto 1\cdot e^{-iE_1t}-V_1V_2e^{-iE_2\tau}e^{-iE_1(t-\tau)}-   \\
            {}-V_1V_3e^{-iE_2(T+\tau)}e^{-iE_1(t-\tau-T)}-   \\
            {}-V_2V_3e^{-iE_1\tau}e^{-iE_2T}e^{-iE_1(t-\tau-T)},
\end{multline}
\begin{multline}\label{C2after3}
    C_2 \propto -iV_1e^{-iE_2t}-iV_2e^{-iE_1\tau}e^{-iE_2(t-\tau)}-   \\
            {}-iV_3e^{-iE_1(T+\tau)}e^{-iE_2(t-\tau-T)}.
\end{multline}

The complex echo amplitude is proportional to $C_1C_2^*$. Only a product of the second term in Eq.~(\ref{C1after3}) and the third term in Eq.~(\ref{C2after3}) and a product of the third term in Eq.~(\ref{C1after3}) and the second term in Eq.~(\ref{C2after3}) gives us the three-pulse echo signal. As a result of these two contributions the three-pulse echo amplitude from one TLS is proportional to
\begin{equation}
    p_{\rm echo} \propto -iV_1V_2V_3e^{i(E_2-E_1)(t-T-2\tau)}.
\end{equation}
Summing echo amplitudes over all TLS's we have for the total echo amplitude
\begin{equation}
    P_{\rm echo} \propto -i\sum_{\rm TLS} V_1V_2V_3e^{iE(t-T-2\tau)}.
\end{equation}
At the time $t=T+2\tau$ the phase of the echo signal from all TLS's becomes equal to zero so we have a maximum of the overall echo amplitude $P_{\rm echo} \propto -iV_1V_2V_3N$, where $N$ is the number of resonant TLS's (i.e. TLS's with $\hbar\omega\approx E$).

\section{Diagram Representation of the Echo}

\begin{figure*}
    \centering
    \includegraphics[width=\textwidth]{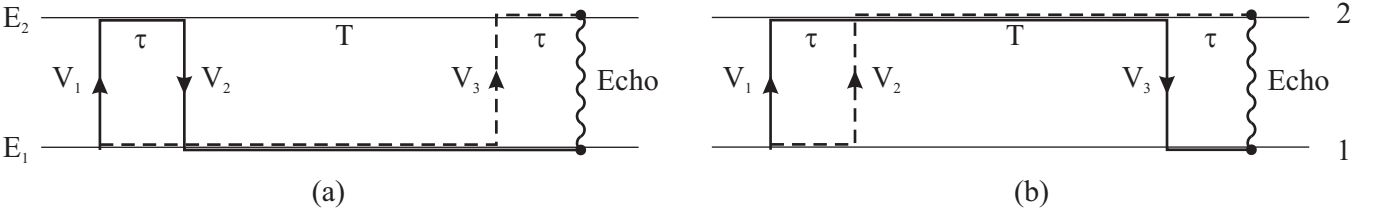}
    \caption{\label{fig:EchoDiagramTLS} Diagrams representation of the three-pulse echo from a TLS. (a)~The first contribution. (b)~The second contribution.}
\end{figure*}

For the future analysis of the echo in magnetic field let us now give a simple diagram representation of the three-pulse echo from one TLS (see Fig.~\ref{fig:EchoDiagramTLS}). For simplicity, we assume that the electric
field amplitudes ${\bf F}_{1,2,3}$ of the three electric pulses
are parallel to each other, i.e. ${\bf F}_i=F_i{\bf e}$, where ${\bf e}$ is a unit polarization vector.

Diagram in Fig.~\ref{fig:EchoDiagramTLS}a corresponds to the product of the second term in Eq.~(\ref{C1after3}) (the full line in the diagram) and the third term in Eq.~(\ref{C2after3}) (the dashed line). Each vertical line (except the final wavy line) corresponds to a transition from one level to another under the action of one of electric pulses. One should ascribe to this line a factor $(-iF_j)\alpha_{kl}$, where $j$ is a number of the pulse, and $\alpha_{kl}$ is a dipole transition matrix element  from the level $l$ to the level $k$ in the TLS. Each horizontal line corresponds to free TLS evolution in the time intervals between or after the pulses.  For example between the neighboring pulses $i$ and $j$ we ascribe to such line a factor $\exp(-iE_k\delta t_{ij})$, where $k$ is a level number, and $\delta t_{ij}$ is a time delay between the $i$ and $j$ pulses. Finally factor $\alpha_{21}$ corresponds to the vertical wavy line. In the final expression all factors corresponding to dashed lines should be taken as complex conjugated \cite{diagram}.

According to these rules we get for the first contribution to the echo amplitude the following expression
\begin{multline}\label{p1}
    p_{\rm echo}^{(1)} \propto (-iF_1)\alpha_{21}\cdot e^{-iE_2\tau}\cdot(-iF_2)\alpha_{12}\cdot e^{-iE_1(t-\tau)}\times \\
        \times\left[e^{-iE_1(\tau+T)}\cdot(-iF_3)\cdot\alpha_{21}\cdot e^{-iE_2(t-T-\tau)} \right]^*\alpha_{21}= \\
            = -iF_1F_2F_3|\alpha_{12}|^2|\alpha_{21}|^2e^{i(E_2-E_1)(t-T-2\tau)}.
\end{multline}
We took here into account that $\alpha_{12}=\alpha_{21}^*$ The second contribution corresponds to the product of the third term in Eq.~(\ref{C1after3}) and the second term in Eq.~(\ref{C2after3})
\begin{multline}\label{p2}
    p_{\rm echo}^{(2)} \propto (-iF_1)\alpha_{21}\cdot e^{-iE_2(\tau+T)}\cdot(-iF_3)\alpha_{12}\times \\
        \times e^{-iE_1(t-\tau-T)}\left[e^{-iE_1\tau}\cdot(-iF_2)\cdot\alpha_{21}\cdot e^{-iE_2(t-\tau)} \right]^*\alpha_{21}= \\
            = -iF_1F_2F_3|\alpha_{12}|^2|\alpha_{21}|^2e^{i(E_2-E_1)(t-T-2\tau)}.
\end{multline}

Comparing Eqs.~(\ref{p1}) and (\ref{p2}) we see that both contributions to the echo amplitude turn out to be identical in this simple case. As a result the total echo amplitude
\begin{equation}
    p_{\rm echo} \propto -iF_1F_2F_3|\alpha_{12}|^2|\alpha_{21}|^2e^{i(E_2-E_1)(t-T-2\tau)}.
\end{equation}

\section{Echo in a Multi-Level System}

\begin{figure*}
    \centering
    \includegraphics[width=\textwidth]{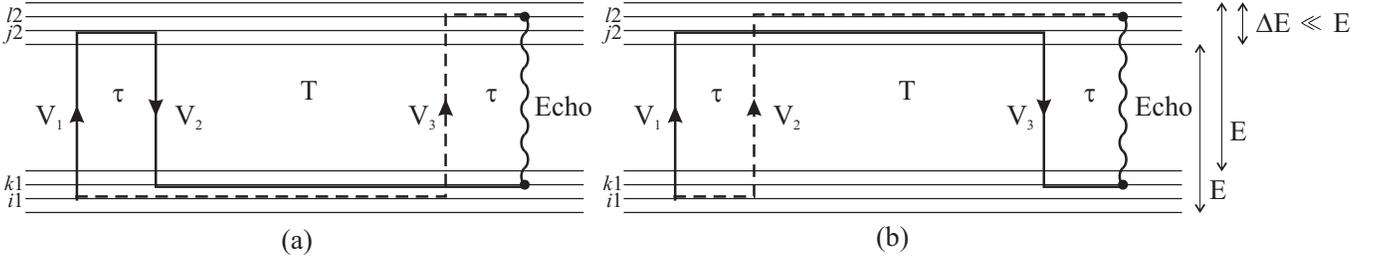}
    \caption{\label{fig:EchoDiagramMult} Diagrams representation of the three-pulse echo in multi-level system. (a)~The first contribution. (b)~The second contribution.}
\end{figure*}

Using these simple diagram rules we can now easily find the contributions to the three-pulse echo signal from a multi-level system. Let us, for example, consider a multi-level system as shown in Fig.~\ref{fig:EchoDiagramMult}. It consists of two identical groups of $\mathcal N$ levels shifted vertically against
each other by some energy $E$ (playing the role of the usual TLS energy). Inside the groups the positions of the levels are arbitrary, i.e. they are not necessarily equidistant. As we will see in the next sections, such a multi-level system describes a TLS with a quadrupole nuclear moments or with dipole-dipole interaction of its nuclear spins.

According to the diagram rules formulated in the previous section we get two contributions to the echo amplitude
    \begin{multline}\label{pi1}
        \pi_{i1,k1,j2,l2}^{(1)}(t)=\frac{1}{\sqrt{\mathcal N}}(-iF_1)\alpha_{j2,i1}\cdot e^{-iE_{j2}\tau}\times \\
        \times(-iF_2)\alpha_{k1,j2}\cdot e^{-iE_{k1}(t-\tau)}\times \\
            \times\left[\frac{1}{\sqrt{\mathcal N}}e^{-iE_{i1}(\tau+T)}\cdot(-iF_3)\alpha_{l2,i1}\cdot e^{-iE_{l2}(t-\tau-T)}\right]^*\alpha_{l2,k1}= \\
            =-\frac{i}{\mathcal N}F_1F_2F_3\alpha_{j2,i1}\alpha_{k1,j2}\alpha_{l2,i1}^*\alpha_{l2,k1}\times \\
                \times\exp[-i(E_{j2}-E_{i1})\tau-i(E_{k1}-E_{i1})T+{} \\
                {}+i(E_{l2}-E_{k1})(t-\tau-T)],
    \end{multline}
and
    \begin{multline}\label{pi2}
        \pi_{i1,k1,j2,l2}^{(2)}(t)=\frac{1}{\sqrt{\mathcal N}}(-iF_1)\alpha_{j2,i1}\cdot e^{-iE_{j2}(\tau+T)}\times \\
        \times(-iF_2)\alpha_{k1,j2}\cdot e^{-iE_{k1}(t-\tau-T)}\times \\
            \times \left[\frac{1}{\sqrt{\mathcal N}}e^{-iE_{i1}\tau}\cdot(-iF_3)\alpha_{l2,i1}\cdot e^{-iE_{l2}(t-\tau)}\right]^*\cdot\alpha_{l2,k1}=\\
            =-\frac{i}{\mathcal N}F_1F_2F_3\alpha_{j2,i1}\alpha_{k1,j2}\alpha_{l2,i1}^*\alpha_{l2,k1}\times \\
                \times\exp[-i(E_{j2}-E_{i1})\tau+i(E_{l2}-E_{j2})T+{} \\
                {}+i(E_{l2}-E_{k1})(t-\tau-T)].
    \end{multline}
The factors $1/\sqrt{\mathcal N}$ in these formulas correspond to the case when the low-energy group levels are equally populated (with a probability $1/{\mathcal N}$) and the high-energy group levels are empty. This is the case for low enough temperatures, $kT\ll E$. On the other hand, to satisfy the previous conditions the temperature must be much larger than the width of the energy splitting in the groups (this width is of the order of the level splitting $\Delta E$), i.e. $kT\gg\Delta E$. The latter inequality corresponds to the usual experimental situation. The two conditions are compatible if $E\gg\Delta E$. In the usual experiments this inequality is met since the resonance frequency $\hbar\omega\approx E\gg \Delta E$. The limitation $E\gg kT$ is not a crucial. The final result can be easily generalized to the case $kT\simeq E$ by including thermal occupation numbers.

Two groups of levels are identical and only shifted by the energy $E$ so we have
\begin{equation}
    E_{l2}=E+E_{l1}, \quad E_{j2}=E+E_{j1}.
\end{equation}
Then we can rewrite Eqs.~(\ref{pi1}) and (\ref{pi2}) as
\begin{multline}
\label{ghvbf}
    \pi_{i1,k1,j2,l2}^{(1)}(t) = -\frac{i}{\mathcal N}F_1F_2F_3\alpha_{ji}^{21}\alpha_{kj}^{12}\alpha_{li}^{*21}\alpha_{lk}^{21}\times \\
        \times\exp[iE(t-2\tau-T)-i(E_{j1}-E_{i1})\tau-{} \\
            {}-i(E_{k1}-E_{i1})T+i(E_{l1}-E_{k1})(t-\tau-T)],
\end{multline}
and
\begin{multline}
\label{cx5fgd}
    \pi_{i1,k1,j2,l2}^{(2)}(t) = -\frac{i}{\mathcal N}F_1F_2F_3\alpha_{ji}^{21}\alpha_{kj}^{12}\alpha_{li}^{*21}\alpha_{lk}^{21}\times \\
        \times\exp[iE(t-2\tau-T)-i(E_{j1}-E_{i1})\tau+{} \\
            {}+i(E_{l1}-E_{j1})T+i(E_{l1}-E_{k1})(t-\tau-T)],
\end{multline}
where we write the indexes $(1, 2)$ of the level groups as superscripts. From Eqs.~(\ref{ghvbf}), (\ref{cx5fgd}) it follows, in the case when $E\gg\Delta E$, that the echo signal appears at $t=2\tau +T$. At this time, summing over all possible combinations of vertical transitions ($i,j,k,l$) between different levels and taking into account that $\alpha_{ij}^{(12)}=\alpha_{ji}^{*(21)}$, we finally get for two contributions to the echo signal from one multi-level system the following expressions
\begin{multline}\label{EchoMlev1}
    p_{\rm echo}^{(1)}(T+2\tau) \propto -\frac{i}{\mathcal N}F_1F_2F_3\times \\
        \times\sum_{i,k}e^{i(E_i-E_k)(\tau+T)} \left|\sum_j\alpha_{ij}^{12}\alpha_{kj}^{*12}e^{iE_j\tau}\right|^2 ,
\end{multline}
and
\begin{multline}\label{EchoMlev2}
    p_{\rm echo}^{(2)}(T+2\tau) \propto -\frac{i}{\mathcal N}F_1F_2F_3\times \\
    \times\sum_{i,k}e^{i(E_i-E_k)\tau} \left|\sum_j\alpha_{ij}^{12}\alpha_{kj}^{*12}e^{iE_j(\tau+T)}\right|^2 .
\end{multline}
We can see that these two contributions are not identical, as opposed to the case of simple TLS.

Now we consider an arbitrary TLS-nuclear spins multi-level system. Like in~\cite{diagram} we use perturbation theory and get finally for the echo amplitude
\begin{multline}\label{EchoMagn}
    P_{\rm echo}(T+2\tau) \propto -iF_1F_2F_3\left(\frac{\Delta_0}{E}\right)^4\times \\
        \times\Bigg[1-\frac{64}{\mathcal N}\sum_{n, m>n}\left(\frac{\Delta}{E}\right)^2\left|(\widetilde{V}_J)_{nm}\right|^2\times \\
            \times\frac{\sin^2(\varepsilon_{nm}\tau/2\hbar)\sin^2(\varepsilon_{nm}(\tau+T)/2\hbar)}
            {\varepsilon_{nm}^2}\Bigg].
\end{multline}
Where $\varepsilon_{mn}=E_m - E_n$ is the distance between the energy levels inside one level group. And $(\widetilde{V}_J)_{nm}$ are transition matrix elements which we define below. This is a general formula for three-pulse echo for any mechanism of TLS-nuclear spin interaction. At $T=0$ and $F_2=F_3$ this formula transforms to the known formula for two-pulse echo amplitude \cite{ShumPar1}. In this sense we can say that the three-pulse echo is a generalization of the two-pulse echo.

\section{Quadrupole Interaction}

\begin{figure*}
    \centering
    \includegraphics[width=0.95\textwidth]{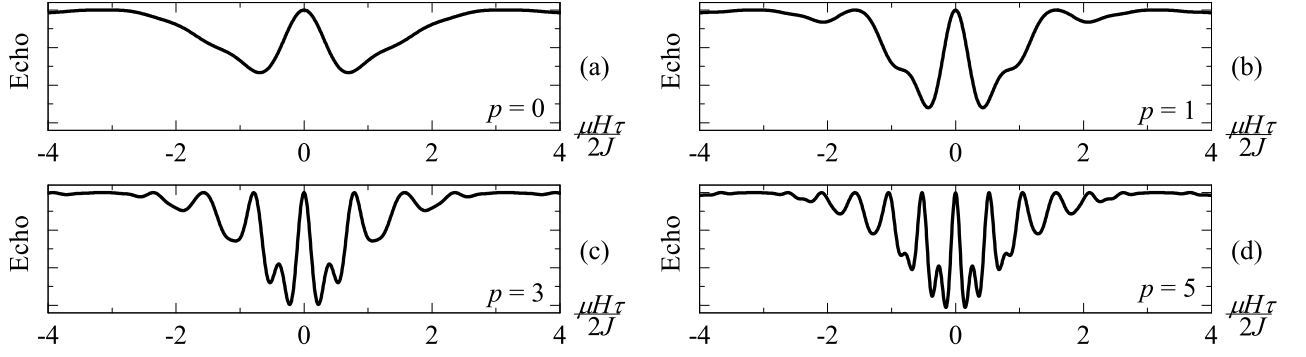}
    \caption{\label{fig:EchoMagn} Echo amplitude as a function of the external magnetic field $\mu H\tau/2J\hbar$ for different ratios $p=T/\tau$.}
\end{figure*}

In this section we specify the general formula, Eq.~(\ref{EchoMagn}), for a TLS with a nuclear quadrupole electric moment in an external magnetic field.

The interaction between the nuclear electric quadrupole moment and gradient of an internal electric field is described by the Hamiltonian~\cite{Abragam}
\begin{equation}
    \widehat{E}_{J}^{(1,2)} = \widehat{Q}_{\alpha\beta}\,\varphi_{\alpha\beta}^{(1,2)}, \quad \varphi_{\alpha\beta}^{(1,2)}\equiv \frac{\partial^2\varphi^{(1,2)}}{\partial r_\alpha\partial r_\beta},
\end{equation}
where $\varphi^{(1,2)}$ is an electrostatic potential at the nuclear site depending on TLS well $(1)$ or $(2)$. The operator of the nuclear quadrupole electric moment is the traceless tensor~\cite{Abragam}
\begin{equation}
    \widehat{Q}_{\alpha\beta} = \frac{eQ}{6J(2J-1)}\left[\frac{3}{2}\left(\hat{J}_\alpha\hat{J}_\beta+\hat{J}_\beta\hat{J}_\alpha\right)-\delta_{\alpha\beta}J(J+1)\right],
\end{equation}
where $\hat{\bf J}$ is the operator of the nuclear spin. Then we define operators
\begin{eqnarray}
    &\widehat{W}_J = \widehat{Q}_{\alpha\beta}\varphi_{\alpha\beta}^{(+)}-{\bf \widehat{M} \cdot H}, \qquad \widehat{V}_J = \widehat{Q}_{\alpha\beta}\varphi^{(-)}_{\alpha\beta},& \nonumber \\
    &\varphi^{(\pm)}_{\alpha\beta} = \frac{\displaystyle \varphi^{(1)}_{\alpha\beta} \pm \varphi^{(2)}_{\alpha\beta}}{\displaystyle 2},& \label{Quad_WsVs}
\end{eqnarray}
where ${\bf H}$ is the external magnetic field, $\widehat{\bf M}$ is a nuclear magnetic dipole moment of the TLS.

Using unitary transformation $\widehat{S}_J$ we can diagonalize matrix $\widehat{W}_J$ and introduce
a new interaction operator $\widehat{\widetilde{V}}_J$
\begin{equation}
\widehat{S}_J\,\widehat{W}_J\,\widehat{S}_J^{-1}\equiv
\widehat{\widetilde{W}}_J, \quad\mbox{and}\quad
\widehat{S}_J\,\widehat{V}_J\,\widehat{S}_J^{-1}\equiv
\widehat{\widetilde{V}}_J .
\label{eq:31}
\end{equation}
Now $\widehat{\widetilde{W}}_J$ is a diagonal matrix in the nuclear spin space ${\cal N}\times {\cal N}$ (for one nuclear spin ${\cal N}=2J+1$). It gives us the nuclear quadrupole energies $E_n$ and fine interlevel spacing $\varepsilon_{mn}=E_m-E_n$ in the average (over the two minima) internal electric field gradient and
in the external magnetic field ${\bf H}$. The unitary transformation $\widehat{S}_J$ can be found in general only by numerical diagonalization of the matrix $\widehat{W}_J$. The transformed matrix $\widehat{\widetilde{V}}_J$ entering the Eq.~(\ref{EchoMagn}) describes the interaction of the TLS with its nuclear quadrupole moment. It is proportional to the difference in electric field gradients in two wells.

Let us further consider the case when Zeeman energy is much larger than the quadrupole level splitting
\begin{equation}
    \widehat{\bf M}\cdot{\bf H} \gg \widehat{Q}_{\alpha\beta}\varphi_{\alpha\beta}^{(+)},
\end{equation}
and distances between the pulses satisfy the condition
\begin{equation}
\label{gyuu}
    \widehat{Q}_{\alpha\beta}\varphi_{\alpha\beta}^{(+)}(T+\tau)/\hbar\lesssim 1,
\end{equation}
In this case we neglect the quadrupole interaction term $\widehat{Q}_{\alpha\beta}\varphi_{\alpha\beta}^{(+)}$ in Eq.~(\ref{Quad_WsVs}) in comparison with Zeeman energy. In a full analogy with two-pulse echo \cite{ShumPar1} we get from Eq.~(\ref{EchoMagn}) the final formula for the echo amplitude in the high magnetic field:
\begin{multline}\label{EchoHighMagn}
    P_{\rm echo} \propto -iV_1V_2V_3\left(\frac{\Delta_0}{E}\right)^4\times \\
        \times\Bigg\{1-C\cdot\bigg[\frac{\sin^2(\mu H(T+\tau)/2J)\sin^2(\mu H \tau/2J)}{H^2}+{}\\
            +\frac{\sin^2(\mu H(T+\tau)/J)\sin^2(\mu H \tau/J)}{4H^2}\bigg]\Bigg\},
\end{multline}
where $C$ is some constant independent of magnetic field.

If we put $T = 0$ in Eq.~(\ref{EchoHighMagn}) we get the existing formula for the two-pulse echo amplitude~\cite{ShumPar1}. To illustrate this formula we plot on Fig.~\ref{fig:EchoMagn} the echo amplitude as a function of the external magnetic field for several values of the ratio $p=T/\tau$. For $T=0$ the magnetic field dependence is the same as for two-pulse echo (Fig.~\ref{fig:EchoMagn}a). But with increasing of the ratio $T/\tau$ the echo amplitude becomes larger (about factor of two) and gets more additional peaks as a function of magnetic field.

As follows from Eq.~(\ref{EchoHighMagn}) the absolute maxima of the echo amplitude take place at
\begin{equation}
\label{AM}
\frac{\mu H\tau}{2J} = \frac{k\pi}{1+p}, \quad k=0,1,2,... \, .
\end{equation}
From positions of these maxima one can calculate magnetic dipole moment $\mu$ of tunneling nuclei. Maxima at
$\mu H\tau/2J = k\pi$ which exist also in a two-pulse echo signal are very shallow and can not be used for this purpose.

\section{Dipole-dipole Interaction}

Let us now consider the case where fine splitting of TLS levels is determined by dipole-dipole interaction of
its nuclear spins
\begin{equation}
\label{HJ}
\widehat{E}_J^{(1,2)} = \sum_i \widehat{\boldsymbol{\mu}}_i {\bf H} + \frac{1}{2}\sum_{ij}
\left[
\frac{\widehat{\boldsymbol{\mu}}_i \widehat{\boldsymbol{\mu}}_j}{r_{ij}^3} - 3\frac{\left(\widehat{\boldsymbol{\mu}}_i {\bf r}_{ij}\right)\left(\widehat{\boldsymbol{\mu}}_j{\bf r}_{ij}\right)}{r_{ij}^5}
\right].
\end{equation}
Here $\widehat{\boldsymbol{\mu}}_i = \mu_i\widehat{{\bf J}}_i/J_i$ magnetic moment operator of $i$th nuclear, $\widehat{\bf J}_i$  is its spin operator, $\bf H$ is the external magnetic field. Radius vector ${\bf r}_{ij} \equiv {\bf r}_{ij}^{(1,2)}$, from $i$th nuclear to $j$th nuclear depends on the well number of TLS (1 or 2). The summation in Eq.~(\ref{HJ}) is taken over all tunneling nuclei.

The spin operators $\widehat{W}_J$ and $\widehat{V}_J$ in this case are symmetric and antisymmetric parts of spin hamiltonian $\widehat{E}_J^{(1,2)}$ correspondingly
\begin{equation}\label{WV}
  \widehat{W}_J = \frac{\widehat{E}_J^{(1)} + \widehat{E}_J^{(2)}}{2} ,\quad
  \widehat{V}_J = \frac{\widehat{E}_J^{(1)} - \widehat{E}_J^{(2)}}{2} .
\end{equation}
Note that the first term of spin Hamiltonian (\ref{HJ}), which is associated with the external magnetic field (the Zeeman part), does not include vectors ${\bf r}_{ij}$ that change under transition of the TLS from one well to another and, therefore, does not contribute to the antisymmetric part of the Hamiltonian $\widehat{V}_J$. The dipole-dipole interaction (the second term of Hamiltonian (\ref{HJ})) enters both operators $\widehat{W}_J$ and $\widehat{V}_J$. Finally we should perform the unitary transformation similar to the previous case, Eq.~{\ref{eq:31}}.

To simplify the problem, let us consider further the limit of high magnetic fields when Zeeman energy is much bigger than the energy of dipole-dipole interaction $E_d$ and $E_d(T+\tau)/\hbar\lesssim 1$ which is similar with (\ref{gyuu}). In addition we restrict ourself the case of dipole-dipole interaction of identical nuclei with spins 1/2 as it takes place in glycerol. Then by analogy with \cite{ShumPar2} we get from general formula, Eq.~(\ref{EchoMagn}), for the echo amplitude the following expression
\begin{multline}
\label{gtq}
    P_{\rm echo} \propto 1-C\Bigg[\frac{\sin^2(\mu H\tau)\sin^2(\mu H(T+\tau))}{(\mu H)^2}+{}\\
        {}+\frac{\sin^2(2\mu H\tau)\sin^2(2\mu H(T+\tau))}{4(\mu H)^2}\Bigg].
\end{multline}
The coefficient $C$ in this expression is now determined by dipole-dipole interaction of TLS nuclear spins $\widehat{V}_J$. We see that Eq.~(\ref{gtq}) is formally identical to Eq.~(\ref{EchoHighMagn}) for the case of quadrupole interaction with $J=1/2$. This coincidence is due to that we have considered a dipole-dipole
interaction of two identical nuclei. It disappears in a more general case.

\section{Conclusions}

We developed a theory of the three-pulse electric-dipole echo in glasses in a magnetic field. As usually two-level systems contribute to the echo signal at low enough temperatures. We have shown that magnetic field dependence of the echo amplitude is due to quadrupole electric moments of tunneling nuclei and/or dipole-dipole interaction of their nuclear spins. In the case when Zeeman energy is bigger than the quadrupole energy or energy of dipole-dipole interaction the magnetic field dependence of the echo amplitude appears to be a simple universal function of magnetic field independently of microscopic structure of TLS's. In comparison with the two-pulse echo the magnetic field dependence has a more reach spectrum of echo oscillations. From positions of the maxima of echo amplitude as a function of magnetic field one can find the magnetic dipole moment of tunneling nuclei.

\section{Acknowledgments}

We are grateful to A.~V.~Shumilin for fruitful discussions and critical reading of the manuscript.
Y.M.B. thanks the St. Petersburg administration for the financial support in 2009, diploma  project
no. 2.4/04-05/103. D.A.P. thanks Alexander von Humboldt-Foundation and Kirchhoff Institute for Physics University of Heidelberg for financial support and hospitality.

\end{document}